\documentstyle[prb,aps,floats,epsf]{revtex}

\begin{document} 
\draft
\sloppy
\raggedbottom 

\twocolumn[\hsize\textwidth\columnwidth\hsize\csname
@twocolumnfalse\endcsname

\title{Electronic Structure of the Chevrel-Phase Compounds 
Sn$_{x}$Mo$_{6}$Se$_{7.5}$: \\ Photoemission Spectroscopy and 
Band-structure Calculations}

\author{K. Kobayashi\cite{presentaddress}}
\address{Department of Physics, University of Tokyo, Hongo 7-3-1, 
Bunkyo-ku, Tokyo 113-0033, Japan}

\author{A. Fujimori} \address{Department of Physics and Department of 
Complexity Science and Engineering,\\ University of Tokyo, Bunkyo-ku, 
Tokyo 113-0033, Japan}

\author{T. Ohtani} \address{Laboratory for Solid State Chemistry, 
Okayama University of Science, Ridai-cho 1-1, Okayama 700-0005, Japan}

\author{I. Dasgupta\cite{presentaddress2}, O. Jepsen, and O. K. Andersen} 
\address{Max-Planck-Institut f{\"u}r Festk{\"o}rperforschung, Postfach 
800665, D-70506, Stuttgart, Germany}

\date{\today}
\maketitle

\begin{abstract}
We have studied the electronic structure of two Chevrel-phase 
compounds, Mo$_6$Se$_{7.5}$ and Sn$_{1.2}$Mo$_6$Se$_{7.5}$, by 
combining photoemission spectroscopy and band-structure calculations.  
Core-level spectra taken with x-ray photoemission spectroscopy show 
systematic core-level shifts, which do not obey a simple rigid-band 
model.  The inverse photoemission spectra imply the existence of an 
energy gap located $\sim 1$ eV above the Fermi level, which is a 
characteristic feature of the electronic structure of the Chevrel 
compounds.  Quantitative comparison between the photoemission spectra 
and the band-structure calculations have been made.  While good 
agreement between theory and experiment in the wide energy range was 
obtained as already reported in previous studies, we found that the 
high density of states near the Fermi level predicted theoretically 
due to the Van Hove singularity is considerably reduced in the 
experimental spectra taken with higher energy resolution than in the 
previous reports.  Possible origins are proposed to explain this 
observation.
\end{abstract}


\vskip1pc]
\narrowtext

\section{Introduction}
The Chevrel-phase compounds, which have been extensively studied in 
1970's as one of the largest families of 
superconductors~\cite{STC,STM,FischerAP78}, have recently invoked 
renewed interest as a candidate for new thermoelectric 
materials~\cite{RochePRB1999,NunesPRB1999}.  Their general formula is 
\textit{A}$_x$Mo$_6$\textit{X}$_8$, where \textit{X} stands for a 
chalcogen atom, namely, S, Se, or Te and \textit{A} atom is an alkali 
metal, alkaline earth, simple metal, transition metal, noble metal, or 
rare earth~\cite{STM}.  Up to now, dozens of compounds with this 
formula are known, a large portion of which show superconductivity.  
Crystallographically, they have a rather remarkable structure 
consisting of Mo$_6$\textit{X}$_8$ clusters.  In this sense, the 
Chevrel system is a forerunner of cluster-based superconductors like 
the fullerenes.  Such materials tend to have rather exotic electronic 
structures, which gives us a unique opportunity to search for new 
compounds with novel properties.

Theoretically, a number of band-structure calculations have been 
reported for the Chevrel-phase 
compounds~\cite{MattheissPRB77,BullettPRL77,AndersenPRB78,
JarlborgPRL80,STC-I-Nohl}.  Two characteristic features should be 
remarked.  The first is the high density of states (DOS) near the 
Fermi level ($E_F$) mainly due to flat bands of Mo 4\textit{d} 
character, which would favor a high superconducting transition 
temperature according to standard BCS theory.  In fact, the Fermi 
level is located close to a Van Hove singularity 
(VHS)~\cite{AndersenPRB78}.  In addition, the flat bands mean a low 
Fermi velocity, which leads to a short coherence length and hence a 
high critical magnetic field ($H_{C2}$)~\cite{FischerAP78}.  Second, 
there exists an energy-gap-like structure about 1 eV above $E_F$.  
This arises from a splitting between the bonding and anti-bonding 
states of the Mo 4\textit{d} manifold and its position relative to 
$E_F$ depends on the number of electrons in the cluster.  It is this 
feature as well as the cluster-based crystal structure that gives some 
researchers the hope that there may exist new thermoelectric materials 
among the Chevrel-phase compounds~\cite{RochePRB1999,NunesPRB1999}.

Experimentally, photoemission spectroscopy (PES) is one of the most 
useful methods to investigate the electronic structures, and it has 
already been applied to this system several times.  Following the 
earlier work by Ihara and Kimura~\cite{IharaJJAP78}, many studies have 
been reported including x-ray photoemission spectroscopy (XPS), 
ultraviolet photoemission spectroscopy 
(UPS)~\cite{KurmaevSSC81,FujimoriPRB86,BrownPRB86,SugaJPSJ86,%
YashonathSSC81,NamatameJJAP89}, and x-ray emission 
spectroscopy~\cite{KurmaevSSC81}.  Although the measured valence-band 
spectra were compared with band-structure calculations and good 
agreement was obtained between experiment and 
theory~\cite{KurmaevSSC81,FujimoriPRB86}, the experiment has not been 
performed using high enough resolution in order to access the 
electronic structure near $E_F$.  The information about the unoccupied 
states to be measured by inverse photoemission spectroscopy (IPES) has 
also been lacking.

In this paper, we will report on a study of the electronic structure 
of two Chevrel-phase compounds, Mo$_6$Se$_{7.5}$ and 
Sn$_{1.2}$Mo$_6$Se$_{7.5}$, combining the results of PES, IPES, and 
the band-structure calculations.  First, the experimental results of 
the core-level, valence-band, and conduction-band spectra will be 
shown.  The valence-band spectra were taken with much higher 
resolution than in previous reports.  Second, after the results of the 
band-structure calculations for Mo$_6$Se$_{8}$ and SnMo$_6$Se$_{8}$ 
are reported, quantitative comparison between the experimental PES and 
IPES spectra and the theoretical spectra derived from the 
band-structure calculations will be performed.  We finally discuss to 
which extent the theory explains the experimental results.

\section{Experiment}
\subsection{Sample Preparation}
Polycrystalline samples of Mo$_6$Se$_{7.5}$ and 
Sn$_{1.2}$Mo$_6$Se$_{7.5}$ were prepared as follows.  For 
Mo$_6$Se$_{7.5}$, a mixture of Mo and Se with the desired ratio was 
sealed in an evacuated silica tube, and was then heated from 
200$^\circ$C to 900$^\circ$C in a rate of 100$^\circ$C/hour, followed 
by annealing at 900$^\circ$C for 12 hours.  The product was ground and 
pressed into a pellet, and was then annealed at 1200$^\circ$C for 
three days.  For Sn$_{1.2}$Mo$_6$Se$_{7.5}$, a mixture of the desired 
ratio of Sn, Mo and Se powders was heated in an evacuated silica tube 
at 200$^\circ$C for 12 hours and 250$^\circ$C for 12 hours, and was 
then heated up to 800$^\circ$C in a rate of 100$^\circ$C/hour, 
followed by annealing at 800$^\circ$C for 24 hours.  The product was 
pressed into a pellet and was annealed again at 1000$^\circ$C for a 
week.  X-ray diffraction patterns of both samples were successfully 
indexed on the basis of the Chevrel structure.  The hexagonal lattice 
parameters were determined to be $a = 9.568$ {\AA} and 9.521 {\AA} and 
$c = 11.180$ {\AA} and 11.838 {\AA} for Mo$_6$Se$_{7.5}$ and 
Sn$_{1.2}$Mo$_6$Se$_{7.5}$, respectively.  The Sn$_x$Mo$_6$Se$_8$ 
phases were stabilized when the atomic ratio between Mo and Se was 
slightly non-stoichiometric.  While the x-ray diffraction does not 
show any extra phase caused by the non-stoichiometry, it is not known 
whether Se deficiency or excess Mo is responsible for this 
non-stoichiometry~\cite{FlukigerSTM}.  Judging from the result for the 
sulfide Chevrel-phase compounds~\cite{WadaJLC1985}, however, the 
former is more likely rather than the latter.  Transitions to the 
superconducting states were found to occur at 5--8 K for 
Mo$_6$Se$_{7.5}$ and 2--8 K for Sn$_{1.2}$Mo$_6$Se$_{7.5}$, being 
consistent with previous reports.

\subsection{Photoemission Measurements}
The XPS measurements were done using the Mg K$\alpha$ line 
($h\nu=1253.6$ eV) and photoelectrons were collected using a PHI 
double-pass cylindrical-mirror analyzer.  The UPS measurements were 
made using the He~I and He~II resonance lines ($h\nu =$ 21.2 eV and 
40.8 eV, respectively) and a VSW hemi-spherical analyzer.  The IPES or 
Bremsstrahlung-isochromat spectroscopy (BIS) measurements were 
performed by detecting photons of \(h\nu=\) 1486.6 eV using a quartz 
monochromator.  The XPS and BIS measurements were made at 
liquid-nitrogen temperature, and the UPS measurements at $\sim 28$ K. 
We did the energy calibration and the estimation of the instrumental 
resolution by using Au evaporated on the surface of the samples after 
each measurement.  They were performed for XPS by defining Au 
$4f_{7/2} = $ 84.0 eV, and for UPS and BIS by measuring the Fermi 
edge.  The total resolution was $\sim 1$ eV, $\sim 35$ meV, $\sim 80$ 
meV, and $\sim 1$ eV for XPS, He~I UPS, He~II UPS, and BIS, 
respectively.

The samples were scraped \textit{in situ} with a diamond file in every 
measurement.  During the XPS measurements, the intensity of the O 
1\textit{s} core-level signal, which indicates contaminations on the 
sample surfaces, did not increase for several hours, once it had been 
almost removed.  Therefore, the measurements of XPS and BIS were 
undertaken by scraping the samples every several hours.  Scraping was, 
however, done more frequently for the UPS measurements because UPS is 
more surface-sensitive than XPS and BIS.

\section{Experimental Results}
\subsection{Core-level Spectra}
\begin{figure}[htb!]
\center \epsfxsize=84mm \epsfbox{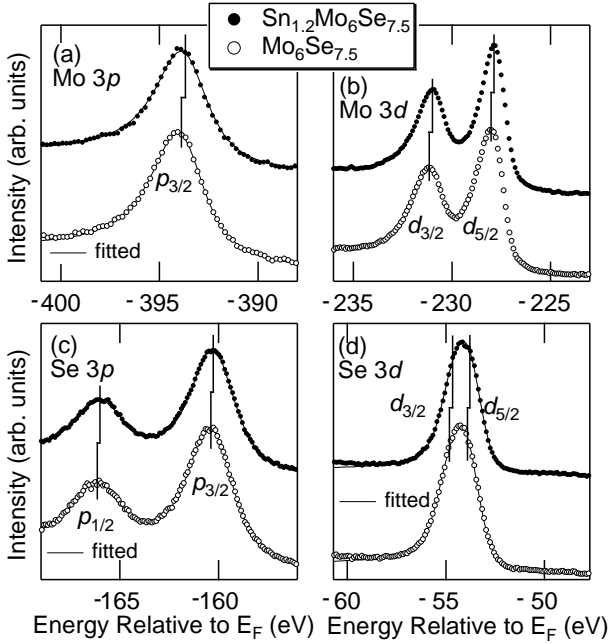} \caption{Mo 
3\textit{p} (a), Mo 3\textit{d} (b), Se 3\textit{p} (c), and Se 
3\textit{d} (d) core-level spectra of Sn$_{1.2}$Mo$_6$Se$_{7.5}$ and 
Mo$_6$Se$_{7.5}$ by closed and open circles, respectively.  In (a), 
(c), and (d), the results of the lineshape fitting are also shown by 
solid curves with vertical bars indicating the peak positions obtained 
through the lineshape analysis.  Only in (b), the positions of the 
intensity maxima are indicated with the vertical bars since the Mo 
3\textit{d} spectra contain signals from the Se 3\textit{s} core 
level.  The obtained peak positions are listed in 
Table~\ref{core_result}.}
\label{XPScore} 
\end{figure}

Figures~\ref{XPScore} (a), (b), (c), and (d) show the Mo 3\textit{p}, 
Mo 3\textit{d}, Se 3\textit{p}, and Se 3\textit{d} core-level spectra 
of Mo$_6$Se$_{7.5}$ and Sn$_{1.2}$Mo$_6$Se$_{7.5}$ obtained by XPS, 
respectively.  The horizontal axis ($E$) measures the energy relative 
to $E_{F}$ and the binding energy ($E_{B}$) is given by $-E$.  Two 
observations are worth mentioning.  First, systematic core-level 
shifts occur in going from Mo$_6$Se$_{7.5}$ to 
Sn$_{1.2}$Mo$_6$Se$_{7.5}$.  Second, the lineshape of the Mo core 
level looks more asymmetric with longer tail towards higher binding 
energy than that of the Se core level.  This is a feature common in 
both compounds.

To be more quantitative on those points, we made a lineshape analysis 
by means of the least-square fitting.  It is assumed that each 
core-level peak has Mahan's asymmetric lineshape reflecting the effect 
of core-hole screening by conduction electrons in 
metals~\cite{HufnerP-S}.  The degree of asymmetry is characterized by 
the singularity index $\alpha$ given by
\begin{equation}
  \alpha = 2 \sum_l (2l+1) (\frac{\delta_l}{\pi})^2 = \sum_l
  \frac{{q_l}^2}{2(2l+1)},
\end{equation}
where $q_l$ is the charge of the conduction electrons with angular 
momentum $l$ which screens the core hole and $\delta_l$ is the phase 
shift, satisfying Friedel's sum rule $\sum_l q_l = 1$.  The lineshape 
is convoluted with a Gaussian and a Lorentzian function which 
represent the instrumental resolution and the core-hole lifetime 
broadening, respectively.  The integral background~\cite{HufnerP-S} is 
also assumed.  The lineshape analysis could be successfully made 
except for the Mo 3\textit{d} core-level spectra which contain the 
weak Se 3\textit{s} core-level peak at $E\sim -229$ eV. The results 
are shown for the Mo 3\textit{p}, Se 3\textit{p}, and Se 3\textit{d} 
core-level spectra of both compounds in Figs.~\ref{XPScore} (a), (c), 
and (d) by solid curves, respectively.  The fitted parameters are 
shown in Table~\ref{core_result}.

\begin{table}[htb!]
\caption{Core-level peak position $E$ and singularity index $\alpha$ 
obtained by the lineshape analysis.  Only for the Mo 3$d$ core level, 
the peak positions determined by their intensity maxima are listed.  
$\Delta E$ denotes the spin-orbit splitting between 3$p_{3/2}$ and 
3$p_{1/2}$ or between 3$d_{5/2}$ and 3$d_{3/2}$.  }
\label{core_result}
\begin{tabular}{c|ccc|ccc}
&\multicolumn{3}{c}{Mo$_6$Se$_{7.5}$}
&\multicolumn{3}{c}{Sn$_{1.2}$Mo$_6$Se$_{7.5}$}\\
Core Level & $E$ (eV) & $\Delta E$ (eV)& $\alpha$ & $E$ (eV)& $\Delta 
E$ (eV) & $\alpha$\\ \hline
{Mo 3$p_{3/2}$} & -393.87 & 17.52 & $\sim 0.18$ & -393.67 & 17.52 & 
$\sim 0.18$\\
{Mo 3$d_{5/2}$} & -228.03 & 3.13 & -- & -227.88 & 3.14 & --\\
{Se 3$p_{3/2}$} & -160.38 & 5.77 & $<0.02$ & -160.27 & 5.77 & $<0.01$\\
{Se 3$d_{5/2}$} & -53.91 & 0.91 & $\sim 0.05$ & -53.78 & 0.88 & 
$\sim 0.07$\\
\end{tabular}
\end{table}

First of all, it is found from Table~\ref{core_result} that the 
singularity indices of the Mo core levels are larger than those of the 
Se core levels, in qualitative agreement with the result of the 
sulfide Chevrel-phase compound~\cite{FujimoriPRB86}.  This observation 
indicates that the contribution from the Mo 4\textit{d} electrons is 
dominant at $E_F$ rather than that from Se 4\textit{p}, being 
qualitatively consistent with the band-structure calculations.  
Secondly, we also found that each core level in Mo$_6$Se$_{7.5}$ is 
located at a binding energy which is 0.1--0.2 eV higher than that in 
Sn$_{1.2}$Mo$_6$Se$_{7.5}$.  As for the trend of the core-level 
shifts, it was reported that the shifts have weak linear correlation 
with the inter-cluster Mo-Mo distance and the rhombohedral lattice 
parameter of the Chevrel systems judging from the core-level and Mo 
1\textit{s} absorption spectra of various $A_x$Mo$_6X_8$ compounds 
(\textit{A} = Pb, Ni, Cu, and so on and \textit{X} = S, Se, and 
Te)~\cite{YashonathSSC81}.  Our results follow the same trend in that 
$E_B$ increases as the Mo-Mo distance increases.  Since the total 
number of electrons in the cluster increases when Sn is added to 
Mo$_6$Se$_{7.5}$, the rigid band model predicts that each core-level 
binding energy should increase in going from Mo$_6$Se$_{7.5}$ to 
Sn$_{1.2}$Mo$_6$Se$_{7.5}$.  The observed core level shifts are 
therefore opposite to those expected from the rigid-band model that Sn 
is an electron donor and that the Fermi level would be raised by 
Sn-doping.  Thus, our results imply that the valence band structure 
itself changes and the system behaves unlike a rigid band model when 
Sn is added interstitially to Mo$_6$Se$_{7.5}$.

\subsection{Valence-band Photoemission Spectra}
\begin{figure}[htb!]
\center \epsfxsize=84mm \epsfbox{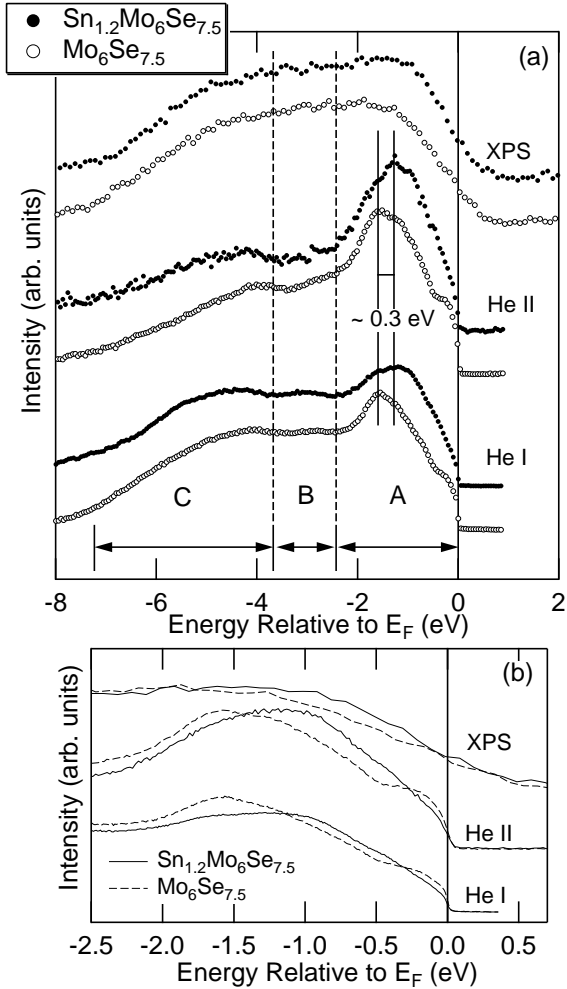} \caption{Valence-band 
photoemission spectra of Mo$_6$Se$_{7.5}$ and 
Sn$_{1.2}$Mo$_6$Se$_{7.5}$ in the entire valence band (a) and in the 
vicinity of $E_{F}$ (b).  The photon energies used for the 
measurements are $h\nu =$ 21.2 eV (He I UPS), 40.8 eV (He II UPS), and 
1253.6 eV (XPS).  All the spectra have been normalized to the area 
between $E = -8$ eV and 0 eV after the background of integral type and 
Henrich type has been subtracted for the XPS and UPS spectra, 
respectively.}
\label{Valence_result} 
\end{figure}

The valence-band XPS and UPS spectra of Mo$_6$Se$_{7.5}$ and 
Sn$_{1.2}$Mo$_6$Se$_{7.5}$ in the entire valence band and in the 
vicinity of $E_{F}$ are shown in Figs.~\ref{Valence_result} (a) and 
(b), respectively.  They have been normalized to the area of the whole 
valence band which spreads between $E \sim -8$ eV and 0 eV after the 
background of integral type~\cite{HufnerP-S} and Henrich 
type~\cite{LiJESRP1993} had been subtracted for the XPS and UPS 
spectra, respectively.

Roughly speaking, three structures are identified in the spectra of 
both compounds.  For the moment, these structures are referred to as A 
(from 0 to $\sim -2$ eV), B ($\sim -3$ eV), and C (from $\sim -4$ to 
$\sim -7$ eV) as indicated in the figure.  Because the Sn 5\textit{s} 
core level is observed at $E \sim -14$ eV, we have to take into 
account only Mo 4\textit{d} and Se 4\textit{p}.  The contribution of 
Mo 5\textit{s} and Sn 5\textit{sp} to the valence-band spectra is 
negligible because of their small numbers of electrons in these 
compounds and of their relatively small cross-sections~\cite{Lindau}.  
Based on the fact that the relative cross-sections of Se 4\textit{p} 
to Mo 4\textit{d} is largest at $h\nu = 1253.6$ eV (XPS) and smallest 
at 40.8 eV (He~II UPS), we can consider A to be of mainly Mo 
4\textit{d} character which shows up as a distinct peak in the He~I 
and He~II UPS spectra.  In a similar way, structures B and C are 
attributed to Se 4\textit{p} character because they appear as strong 
broad features in XPS. In the XPS spectra, it is hard to discriminate 
between B and C because of the low energy resolution.  The above 
assignment is qualitatively consistent with the band-structure 
calculations (see Fig.~\ref{DasguptaDOS}): Mo 4\textit{d} character is 
dominant in the rather narrow energy range from $E =$ 0 to $\sim -2$ 
eV and Se 4\textit{p} character appears as a broad band from $E \sim 
-3$ to $\sim -7$ eV. Quantitative comparison will be made below.

The intensity between $E = -0.3$ and $-1.2$ eV of 
Sn$_{1.2}$Mo$_6$Se$_{7.5}$ is higher than that of Mo$_6$Se$_{7.5}$, 
which holds true for all the three spectra as seen in 
Fig.~\ref{Valence_result} (b).  Here, it may be tempting to consider 
that the Mo 4\textit{d} band of Mo$_6$Se$_{7.5}$ is shifted by $\sim 
0.3$ eV to higher binding energy compared with that of 
Sn$_{1.2}$Mo$_6$Se$_{7.5}$, corresponding to the core-level shift.  We 
should, however, emphasize again that the rigid-band model predicts 
the opposite.  Indeed, the lineshape around $E_{F}$ is qualitatively 
different between both compounds, which clearly means that a simple 
rigid-band model is not applicable to the Chevrel system and that the 
insertion of Sn atoms between the clusters certainly change the 
electronic structure around $E_{F}$.  Alternatively, the shift can be 
partly attributed to the narrowing of the Mo 4\textit{d} band due to 
the increase in the distance between the Mo$_{6}$Se$_{8}$ clusters.  
Generally, when the \textit{X} atom in the Mo$_{6}$$X_{8}$ cluster 
goes from S to Se, the lattice parameters increase due to a larger 
atomic radius of Se.  In a similar way, by inserting large atoms such 
as Sn and Pb, the distance between the clusters 
increases~\cite{FischerAP78}.  This results in the decrease of the Mo 
4\textit{d} bandwidth, and in turn lowers the position of $E_{F}$ 
relative to the other core and valence levels because the Fermi level 
is located close to the top of the Mo 4\textit{d} band in the bonding 
states.  Actually, a slight decrease of the Mo 4\textit{d} band width 
from Mo$_6$Se$_8$ to SnMo$_6$Se$_8$ is also predicted in the 
band-structure calculations as reported below.

\subsection{BIS Spectra}
\begin{figure}[htb!]
\center \epsfxsize=84mm \epsfbox{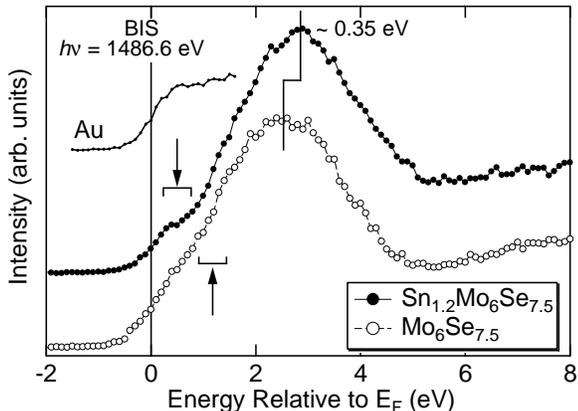} \caption{BIS spectra of 
Sn$_{1.2}$Mo$_6$Se$_{7.5}$ and Mo$_6$Se$_{7.5}$.  The arrows indicate 
the shoulder around 1 eV above $E_F$.  In the top panel, the spectrum 
of gold is also plotted.}
\label{BIS_result} 
\end{figure}

The BIS spectra of Sn$_{1.2}$Mo$_6$Se$_{7.5}$ and Mo$_6$Se$_{7.5}$ are 
shown in Fig.~\ref{BIS_result} with closed and open circles, 
respectively.  The BIS spectrum of Au near the Fermi level is also 
shown in the top panel of the figure for the sake of comparison.  The 
main peak at $E \sim 2$ eV is readily assigned to Mo 4\textit{d} 
character, namely the antibonding states of the Mo 4\textit{d} band, 
because Mo 4\textit{d} has a high density of unoccupied states and its 
cross-section is larger than the other components such as Se 
4\textit{p}.  The broad structure at $E\gtrsim 6$ eV is of Mo 
4\textit{p} character.  As marked in Fig.~\ref{BIS_result}, the main 
peak position of Sn$_{1.2}$Mo$_6$Se$_{7.5}$ is shifted to higher 
energy relative to that of Mo$_6$Se$_{7.5}$ by $\sim 0.35$ eV. This 
shift and its direction are consistent with the PES results, which is 
explained in the same way as above: in going from Mo$_6$Se$_{7.5}$ to 
Sn$_{1.2}$Mo$_6$Se$_{7.5}$, the narrowing of the Mo 4\textit{d} bands 
occurs not only in the bonding state but also in the antibonding state 
and the position of $E_{F}$ is lowered relative to the core levels.  
On the other hand, the centroid of the band in both states should be 
unaffected by the narrowing in the first approximation, making the Mo 
4\textit{d} peak in the unoccupied state shifted away from $E_{F}$ in 
the Sn$_{1.2}$Mo$_6$Se$_{7.5}$ spectra.

Now, a characteristic feature in Fig.~\ref{BIS_result} is the 
existence of a shoulder at $E\sim 0.5$ eV in the 
Sn$_{1.2}$Mo$_6$Se$_{7.5}$ spectrum as marked by an arrow in the 
figure.  This shoulder implies the existence of a dip in the 
unoccupied DOS. Though not so obvious as in the 
Sn$_{1.2}$Mo$_6$Se$_{7.5}$ spectrum, there is a similar shoulder at 
$E\sim 1$ eV in the Mo$_6$Se$_{7.5}$ spectrum.  As mentioned above, 
the band-structure calculations for many Chevrel-phase 
compounds~\cite{MattheissPRB77,BullettPRL77,%
AndersenPRB78,JarlborgPRL80,STC-I-Nohl} have predicted the existence 
of an energy gap around 1 eV above $E_F$.  The observed shoulder is 
consistent with this considering the $\sim 1$ eV energy resolution of 
the BIS measurements [see Fig.~\ref{valdos_result} (a)].  To the best 
of our knowledge, the above observation is the first experimental 
indication of the existence of the gap above $E_F$ in the 
Chevrel-phase compounds.

\section{Band-Structure Calculations}
In order to interpret the experimental results from a theoretical 
viewpoint, we have performed band structure calculations for both 
compounds.  The band structure was calculated self-consistently using 
the local density-functional approximation(LDA) and the scalar 
relativistic linear muffin tin orbital method (LMTO) in the atomic 
sphere approximation (ASA) including the combined correction 
(CC)~\cite{okaPRB75,okaPRL84,ojZP95}.

In the ASA+CC, the one electron potential entering the Schr\"odinger's 
equation is a superposition of overlapping spherical potential wells 
with a position $R$ and radii $s_{R}$, plus a kinetic 
energy error proportional to the fourth power 
of the relative overlap of the spheres. The radii of the 
overlapping muffin-tin spheres are determined by the following 
conditions, that the overlapping muffin-tin potential be the best 
possible approximation to the full potential, and that the error due 
to sphere overlap be acceptable.  For the Chevrel phases with open 
structure, these conditions cannot be achieved only with atom centered 
spheres so several interstitial (empty) spheres were included to 
achieve good sphere packing and an overall good representation of the 
potential.  The radii of the atomic and the interstitial spheres as 
well as the position of the interstitial spheres were calculated by 
using an automatic procedure developed by Krier {\it et 
al}~\cite{krier}.  In the present calculation we allowed an overlap of 
16 percent between atom centered spheres, 18 percent between atom 
centered and the interstitial spheres and 20 percent between the 
interstitial spheres.

The basis set for both the compounds consisted of Mo $5s$, $5p$, $4d$; 
Se $4p$ and the interstitial $s$ LMTO's.  In addition for 
SnMo$_{6}$Se$_{8}$ we included Sn $5s$, $5p$ LMTO's.  Se $s$, $d$; Sn 
$d$ and interstitial $p$-$d$ were downfolded~\cite{okaPRB86}.  This 
treatment not only reduced the size of the secular matrix but also 
avoided distortions of the phase shift of the high partial waves.  
Such distortions or even ghost bands, may occur with the conventional 
LMTO method.  All $k$-space integrations were performed by the 
tetrahedron method~\cite{ojtet} using 254 irreducible $k$-points 
within the Brillouin zone.

\begin{figure}[htb!]
\center \epsfxsize=84mm \epsfbox{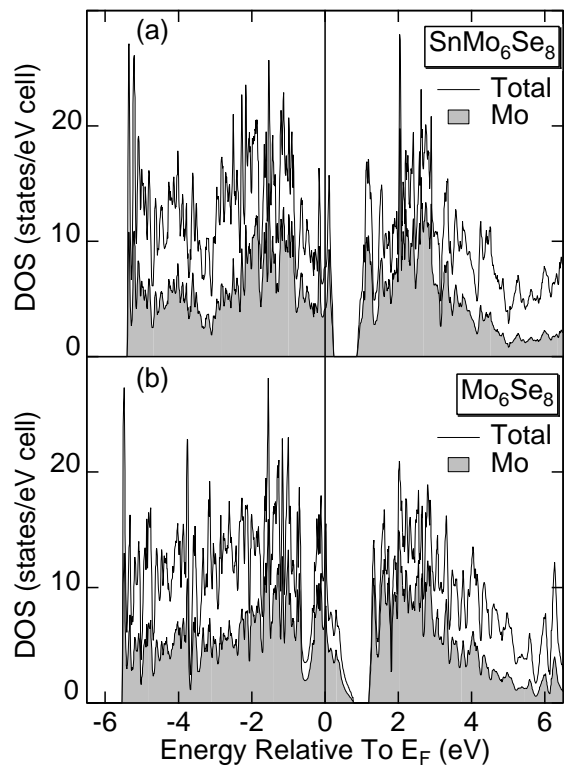} \caption{Calculated 
DOS of (a) SnMo$_6$Se$_8$ and (b) Mo$_6$Se$_8$.  with the total DOS 
and Mo 4\textit{d} partial DOS shown by the solid line and the shaded 
area, respectively.  }
\label{DasguptaDOS} 
\end{figure}

The calculated density of states (DOS) are shown in 
Figs.~\ref{DasguptaDOS} (a) and (b) for SnMo$_6$Se$_{8}$ and 
Mo$_6$Se$_{8}$, respectively, and they are consistent with the 
previous reports on other Chevrel-phase 
compounds~\cite{MattheissPRB77,BullettPRL77,AndersenPRB78,
JarlborgPRL80,STC-I-Nohl}.  The DOS at $E_F$ is very high and situated 
near the VHS, which explains the asymmetric lineshape of the Mo 
core-level spectra.  The width of the Mo 4\textit{d} band in the 
bonding states decreases from 6.4 eV to 5.7 eV in going from 
Mo$_6$Se$_{8}$ to SnMo$_6$Se$_{8}$, which may be partly responsible 
for the aforementioned non-rigid-band-like behavior.  It should be 
noted here that the role of the insertion of Sn atoms between the 
clusters is to change not only the number of electrons but also the 
cluster-cluster interactions.

\section{Comparison between theory and experiment}
In this section, we will make a quantitative comparison between the 
photoemission spectra and theoretical spectra derived from the 
band-structure calculations.  First, it is necessary to re-define the 
Fermi level of the calculation, because due to the Se deficiency the 
Fermi level of Sn$_{1.2}$Mo$_6$Se$_{7.5}$ (Mo$_6$Se$_{7.5}$) is 
apparently higher than SnMo$_6$Se$_8$ (Mo$_6$Se$_8$) by 1.4 (1.0) 
electrons.  The shift corresponding to this in the band-structure 
calculation is 0.072 eV and 0.085 eV for Sn$_{1.2}$Mo$_6$Se$_{7.5}$ 
and Mo$_6$Se$_{7.5}$, respectively.  To derive the theoretical 
photoemission spectra from the results shown in Fig.~\ref{DasguptaDOS} 
we took into account the contribution of Mo 4\textit{d} and Se 
4\textit{p} only.  The partial DOS of both components has been 
weighted by the corresponding photoionization cross-sections at each 
photon energy, and this weighted DOS has been broadened by convoluting 
with a Gaussian and a Lorentzian which represent the instrumental 
resolution and the lifetime broadening, respectively.  We have assumed 
that the lifetime width is linear in energy \textit{E} measured from 
$E_F$, {\it i.e.} FWHM $w=\alpha\vert E - E_F \vert$.  The coefficient 
$\alpha$, which phenomenologically represents the intensity of the 
lifetime of the photo-hole with increasing binding energy, is a 
parameter which is determined so that the measured spectra are well 
reproduced.  For both compounds we have taken $\alpha = 0.24$ and 0.40 
for the photoemission and inverse-photoemission spectra, respectively.  
The spectra have been normalized to their total area.

\begin{figure}[htb]
\center \epsfxsize=84mm \epsfbox{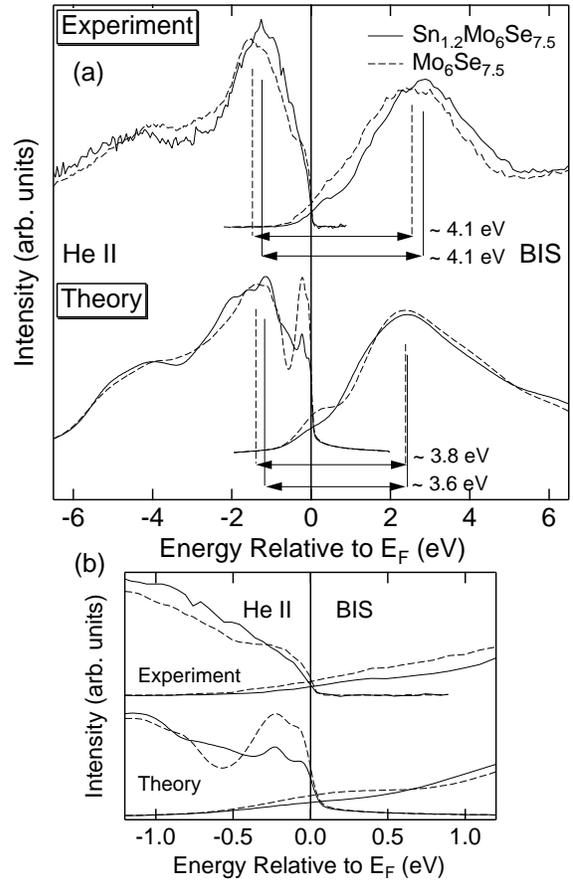} \caption{ (a) Comparison 
between theory and experiment for the He II and BIS spectra of 
Sn$_{1.2}$Mo$_6$Se$_{7.5}$ with solid curves and Mo$_6$Se$_{7.5}$ with 
dashed curves in the wide energy range.  The solid and dashed vertical 
bars indicate the position of the intensity maximum of the main 
structure of the Sn$_{1.2}$Mo$_6$Se$_{7.5}$ and Mo$_6$Se$_{7.5}$ 
spectra, respectively.  The energy splitting between the bonding and 
anti-bonding states of the Mo 4\textit{d} band are also shown.  (b) 
The enlarged plots of (a) around the vicinity of $E_{F}$.}
\label{valdos_result}
\end{figure}

As shown in Fig.~\ref{valdos_result} (a), the agreement in the wide 
energy range between theory and experiment is quite satisfactory as 
has been already reported~\cite{KurmaevSSC81,FujimoriPRB86}.  The main 
features observed around $E=2$--3 eV, $-(1$--$2)$ eV, and $-(3$--$6)$ 
eV in the BIS and PES spectra are well reproduced in the theoretical 
simulation.  As shown in the figure, the energy difference between the 
main peak in the He II spectra and that in the BIS spectra equals 
$\sim 4.1$ eV for both compounds, while the counterpart in the theory 
is $\sim 3.8$ eV for Mo$_6$Se$_{7.5}$ and $\sim 3.6$ eV for 
Sn$_{1.2}$Mo$_6$Se$_{7.5}$.  Although the theoretical values are 
slightly smaller than the experimental ones, the values of the energy 
splittings, which directly reflects the electronic structure of the 
Mo$_6$Se$_{8}$ cluster, fall in the same range for both theory and 
experiment.  In the theoretical studies, it is known that a 
molecular-cluster approach is a good first 
approximation~\cite{JarlborgPRL80}, which is due to the localized 
nature of the Mo 4\textit{d} electrons within the cluster.

In spite of that agreement, however, there exists a large discrepancy 
just around $E_{F}$ between the experimental and theoretical He II 
spectra as shown in Fig.~\ref{valdos_result} (b): while the intensity 
of the Mo$_6$Se$_{7.5}$ spectra is larger than that of the 
Sn$_{1.2}$Mo$_6$Se$_{7.5}$ spectra, which qualitatively holds true in 
theory and experiment, the absolute intensity is not reproduced at 
all.  Indeed, the sharp peak in the theoretical spectra originating 
from the VHS completely disappears in the experimental spectra.  
Although the energy resolution is not high enough to check this point 
for the BIS spectra, one can see a similar tendency in the BIS 
spectra, too.  We believe that the disappearance of the VHS is 
intrinsic judging from the overall good agreement between theory and 
experiment in the wide energy range.  Actually, there are several 
effects which have been neglected in the band-structure calculations.  
The remarkable reduction of the PES intensity near $E_{F}$ indicates 
that such an effect plays a significant role in the spectra on the low 
energy scale.  Because the total spectral weight should be conserved 
if integrated to a sufficiently high energy, the significant weight 
near $E_{F}$ which is theoretically predicted is supposed to be 
transferred to the region away from $E_{F}$ (at least 1--2 eV from 
$E_{F}$) in the experimental spectra.  Similar discrepancy between 
theory and experiment exists in the photoemission spectra of another 
cluster-based superconductor K$_{3}$C$_{60}$~\cite{MorikawaSSC1993}.

Several candidates which can cause the above phenomenon may be listed.  
Firstly, disorder due to the non-stoichiometry.  The Se deficiency 
(and the excess of Sn for Sn$_{1.2}$Mo$_6$Se$_{7.5}$) may create a 
random potential which reduces the coherence of the electrons, leading 
to the decrease of the spectral intensity near $E_{F}$.  It has been 
reported that static disorder causes such an effect in the PES spectra 
of TaSe$_{2}$~\cite{ZwickPRL1998}.  A PES study for more 
stoichiometric single crystals of Chevrel-phase compounds will clarify 
this point.  Secondly, we may consider electron-phonon interactions.  
It was reported that the electron-phonon coupling constant ($\lambda$) 
in the Chevrel-phase compound is as large as $\sim 1$ 
(Ref.~\onlinecite{FischerAP78}), a value which may reduce the spectral 
intensity to almost half of the theoretical 
value~\cite{EngelsbergPR1963}.  The same explanation has been made for 
K$_{3}$C$_{60}$~\cite{GunnarsonPRL1995}.  The typical energy of 
phonons in the Chevrel-phase compounds is, however, only $\sim 10$ meV 
(Ref.~\onlinecite{FischerAP78}), which cannot cause the transfer of 
spectral weight of the order of $\gtrsim 1$ eV. The third candidate is 
the electron-electron interaction, which is presumably significant in 
the Mo$_6$Se$_{8}$ clusters containing localized 4\textit{d} 
electrons.  In 3\textit{d}-transition-metal oxides with strong 
electron correlation, for example, it has been reported that the large 
spectral weight around $E_{F}$ is transferred to the higher biding 
energy farther from $E_{F}$ as an ``incoherent'' 
part~\cite{InouePRL95,MorikawaPRB95}.  While not only one but a 
combination of these three effects may thoroughly explain the 
discrepancy between theory and experiment, the electron-electron 
interaction is the most likely responsible among them, because the 
transfer of spectral weight of the order of $\gtrsim 1$ eV can be only 
explained by taking this interaction into account.  Actually, the 
significance of electron correlation effects in the Mo 4\textit{d} 
band in the Chevrel cluster were previously pointed out by Brusetti 
\textit{et al}~\cite{BrusettiPRB1995}.  Finally, it should be remarked 
that the vanishing of the high DOS due to the VHS is not peculiar to 
this system but has been observed in several superconductors such as 
the boro-carbides and the A15 compounds~\cite{FujimoriPRB1994}.  The 
reason for this universal observations remains to be clarified in the 
future.

\section{Conclusion}
We have studied the electronic structure of two Chevrel-phase 
compounds, Mo$_6$Se$_{7.5}$ and Sn$_{1.2}$Mo$_6$Se$_{7.5}$, using PES 
experiment and band-structure calculations.  The XPS core-level 
spectra have revealed systematic shifts, for which the change of the 
Mo-Mo inter-cluster distances may be responsible.  From the fact that 
the valence-band spectra do not agree with the rigid-band model, we 
propose that the narrowing of the Mo 4\textit{d} bands caused by the 
insertion of Sn atoms explains the observed shift.  An indication of 
the energy gap located $\sim 1$ eV above the Fermi level 
characteristic of the Chevrel system was obtained in the BIS spectra.  
We have also calculated the band-structure of Mo$_6$Se$_{8}$ and 
SnMo$_6$Se$_{8}$ and compared them with the experiment.  The overall 
good agreement between theory and experiment in the wide energy range 
shows that the LDA is valid for these compounds.  On the other hand, 
the high DOS due to the VHS was reduced in the experimental spectra.  
While it has been reported several times that band-structure 
calculations reproduce well the experimental valence band 
spectra~\cite{KurmaevSSC81,FujimoriPRB86}, such a discrepancy was 
first found in the present study because of higher energy resolution 
than the previous work.

Our results imply that the insertion of various atoms between the 
clusters influences the electronic structure around $E_{F}$ through a 
change in the inter-cluster interaction resulting in the change beyond 
the simple rigid-band picture.  The flexibility of the insertion of 
cations into the unique cluster-based structure is attractive and, 
besides the possibility of application as thermoelectric materials, 
there may lie new aspects in the Chevrel-phase and related compounds 
such as one-dimensional compounds made of Chevrel 
clusters~\cite{BrusettiPRB1995}.  Systematic theoretical and 
experimental studies are needed to clarify the relationship between 
the inserted atoms and the electronic structures around $E_{F}$.  Such 
information would be useful to finely tune the electronic properties 
of the Chevrel-phase compounds.

\section*{Acknowledgements} The authors would like to thank T. 
Mizokawa, K. Mamiya, and T. Konishi for their technical supports and 
informative discussions.  One of us (KK) was supported by a Research 
Fellowship of the Japan Society for the Promotion of Science for Young 
Scientists.

\end{document}